\documentclass[prb,twocolumn,showpacs,superscriptaddress,preprintnumbers,amssymb]{revtex4}
\usepackage{graphicx}
\usepackage{dcolumn}
\usepackage{bm}
\usepackage{wasysym}
\usepackage{color}

\newcommand{\mcal}{\mathcal}

\newcommand{\beq}{\begin{equation}}
\newcommand{\eeq}{\end{equation}}
\newcommand{\beqn}{\begin{eqnarray}}
\newcommand{\eeqn}{\end{eqnarray}}

\newcommand{\ua}{\uparrow}
\newcommand{\da}{\downarrow}
\newcommand{\ra}{\rightarrow}

\newcommand{\ii}{\mathrm{i}}
\newcommand{\dd}{\mathrm{d}}

\newcommand{\SU}{\mathrm{SU}}
\newcommand{\U}{\mathrm{U}}

\newcommand{\vb}{\mathbf}

\newcommand{\cxb}[1]{\textcolor{black}{#1}}
\newcommand{\cxc}[1]{\textcolor{black}{#1}}
\newcommand{\cxd}[1]{\textcolor{black}{#1}}

\begin{document}

\title{Vortex Fermi Liquid and Strongly Correlated Quantum Bad Metal}

\author{Nayan Myerson-Jain}

\affiliation{Department of Physics, University of California,
Santa Barbara, CA 93106}

\author{Chao-Ming Jian}
\affiliation{Department of Physics, Cornell University, Ithaca,
New York 14853, USA}

\author{Cenke Xu}

\affiliation{Department of Physics, University of California,
Santa Barbara, CA 93106}

\begin{abstract}

The semiclassical description of two-dimensional ($2d$) metals
based on the quasiparticle picture suggests that there is a
universal threshold of the resistivity: the resistivity of a $2d$
metal is bounded by the so called Mott-Ioffe-Regal (MIR) limit,
which is at the order of $h/e^2$. If a system remains metallic
while its resistivity is beyond the MIR limit, it is referred to
as a ``bad metal", which challenges our theoretical understanding
as the very notion of quasiparticles is invalidated. The
description of the system becomes even more challenging when there
is also strong correlation between the electrons. Partly motivated
by the recent experiment on transition metal dichalcogenides
moir\'{e} heterostructure, we seek for understanding of strongly
correlated bad metals whose resistivity far exceeds the MIR limit.
For some strongly correlated bad metals, though a microscopic
description based on electron quasiparticles fails, a tractable
dual description based on the ``vortex of charge" is still
possible. We construct a concrete example of such strongly
correlated bad metals where vortices are fermions with a Fermi
surface, and we demonstrate that its resistivity can be
exceptionally large at zero temperature. And when extra charge
$\delta n_e$ is doped into the system away from half-filling, a
small Drude weight proportional to $(\delta n_e)^2$ will emerge in
the optical conductivity .

\end{abstract}

\date{\today}

\maketitle

\section{Introduction}

The most rudimentary description of a metal relies on the notion
of quasiparticles, $i.e.$ an electron near the Fermi surface can
be well approximated as a wave packet between two consecutive
elastic scatterings with impurities. This picture requires that
$l_\mathrm{m}k_F > 1$, where $l_\mathrm{m}$ is the mean free path
of the electrons from scattering with the
impurities~\cite{bad_metal}. When $l_\mathrm{m}k_F \sim 1$, the
resistivity of a two dimensional system is of the order of
$h/e^2$, which is also known as the Mott-Ioffe-Regal (MIR) limit.
The common wisdom is that, for noninteracting electrons, when the
resistivity of a $2d$ metal exceeds the MIR limit, not only would
the rudimentary description of the system fail, the system would
actually become an insulator due to the Anderson localization.
The potential metal-insulator transition (MIT) of a noninteracting
$2d$ electron system~\cite{andersonreview} (within certain
symmetry class such as the symplectic) should happen when the
resistivity is of order $h/e^2$.

In strongly interacting electron systems, the universal threshold
of resistivity $h/e^2$ still appears to hold. For electrons at
half-filling (on average one electron per site) on a lattice, the
competition between the interaction and kinetic energy can lead to
an interaction-driven MIT between a metal and a Mott insulator
phase. When the insulator is a particular type of spin liquid
phase, this MIT can be understood through a parton
construction~\cite{lee2005,senthilMIT}, and the total resistivity
follows the Ioffe-Larkin rule~\cite{IoffeLarkin} $\rho = \rho_b +
\rho_f$, where $\rho_b$ and $\rho_f$ are resistivity from the
bosonic and fermionic partons respectively. $\rho_f$ is a smooth
function across the MIT, and at low temperature $\rho_f$ mostly
arises from disorder, which is expected to be small if we assume
weak disorder. Hence at the MIT, the critical resistivity is
mostly dominated by the bosonic parton $\rho_b$. The critical
resistivity $\rho_b$ is expected to be of order $h/e^2$ (though in
the DC limit $\rho_b$ may acquire an extra factor of $7 \sim 8$,
based on analytical evaluation in certain theoretical
limit~\cite{resistivity2})~\footnote{Here we also note that a weak
disorder, or Umklapp process is needed to remove the logarithmic
divergence of conductivity due to thermal fluctuation, which is
predicted by hydrodynamics~\cite{delahydro,kovtunreview}.}.


Hence in most noninteracting as well as strongly interacting
systems that we have understood, the resistivity of a $2d$
metallic system should be roughly bounded by the MIR limit. Hence
if a $2d$ system remains metallic while its resistivity far
exceeds the MIR limit, it challenges our theoretical
understanding. These exotic metals are referred to as ``bad
metals"~\cite{bad_metal}.
The recent experiment on transition metal dichalcogenides (TMD)
revealed the existence of a novel interaction-driven
MIT~\cite{tmdmit1}, where the universal MIR limit is violated: the
DC critical resistivity at the MIT exceeds the MIR limit by nearly
two orders of magnitude. The system is supposedly modelled by an
extended Hubbard model of spin-1/2 electrons on a triangular
moir\'{e} lattice~\cite{tmdhubbard,tmdhubbard2}, but the
experimental finding is qualitatively beyond the previous theory
of MIT. A few recent theoretical
proposals~\cite{fractionalMIT,senthilmit3} were made in order to
understand this exotic MIT.
The experiment mentioned above only revealed a critical point
whose resistivity is clearly beyond the MIR limit. Given the
current experimental finding and the strongly interacting nature
of the system, it is natural to ask, {\it can there also be a
stable bad metal phase of strongly correlated electrons, whose
properties can be evaluated in a tractable way?}

In this work we discuss the construction of a {\it quantum} bad
metal state {\it with longitudinal transport only}; the electrical
resistivity $\rho_e$ of the state can far exceed the MIR limit
even with weak disorder, at zero and low temperature. \cxc{It is
worth noting that the phenomenology of the state we construct is
different from the original example of ``bad metal" (hole doped
cuprates) discussed in Ref.~\onlinecite{bad_metal}, where the
resistivity increases with temperature monotonically and exceeds
the MIR limit at high temperature; while the resistivity of our
``quantum bad metal" remains finite and large {\it at zero
temperature}, and clearly violates the MIR bound.} Our
construction is formulated through the dual degrees of freedom of
``charge vortex". The particle-vortex duality has a long
history~\cite{peskindual,halperindual,leedual}. This duality was
originally discussed for bosons, but recent developments have
generalized the duality to fermion-vortex
duality~\cite{son2015,wangsenthildual,maxashvindual,mrossdual,kachrudual},
as well as Chern-Simons matter theory to free Dirac or Majorana
fermion
duality~\cite{aharony2015,seiberg1,seiberg2,seiberg3,xumajoranadual,raghudual,raghudual2,jiandual}.
And since the particle-vortex duality is still a ``strong-weak"
duality, when the charges are strongly correlated which
invalidates a perturbative description based on quasiparticles,
the vortices are weakly interacting through the dual gauge field,
which facilitates a rudimentary description.

Hence one way to construct a quantum bad metal for charges is to
drive the vortices into a good metal. The vortices can naturally
form a good metal as long as (1) the vortex is a fermion, and (2)
the fermionic vortices form a Fermi surface with a finite density
of states. In the next section we will discuss how exactly a
vortex becomes a fermion in our construction, and how to derive
the charge responses of the system from the physics of vortices.
We would like to clarify that we are not the first to investigate
correlated electrons as vortex liquid. Besides the more well-known
interpretation of composite fermions as ``vortex liquid" in the
context of half-filled Landau level (and similarly for charged
bosons at filling 1)~\cite{HLR,read,son2015,wanghall1,wanghall2};
a metallic phase with {\it anomalously large} conductivity that
emerges in amorphous thin film also motivated discussions of
exotic physics of superconductor
vortices~\cite{vortexliquidfisher,philip,failedsc}. We will
compare our construction with the previous works.

\section{construction of the quantum bad metal}

\subsection{General considerations}

Before we detail our construction, some general considerations can
already be made.

(1) As was pointed out in previous literatures, at least for
charged bosons, the product of the conductivity of the charges and
the conductivity of the vortices is a
constant~\cite{fisherdiffusion,Gazit2015}, $i.e.$ $\sigma_e \sim
1/\sigma_v$. If this relation still (at least approximately) holds
in our construction, it implies that if the vortex conductivity
$\sigma_v$ follows the standard behavior of a good metal at finite
temperature, then the resistivity $\rho_e(T)$ of charge should
decrease with $T$, at least below certain characteristic energy
scale.

(2) A charge vortex can generally be viewed as a point defect with
circulating vorticity of charge current. A charge vortex must
become an anti-vortex under spatial reflection $\mathcal{P}$. This
is because the electric current circulation will reverse its
orientation under reflection. If the vortices form a Fermi
surface, in general it would break $\mathcal{P}$, as a Fermi
surface usually is not invariant under the particle-hole
transformation. The same observation can be made for time-reversal
$\mathcal{T}$: since charge density is invariant under
$\mathcal{T}$, time-reversal would reverse the direction of
electric current circulation. We will discuss later how to
preserve $\mathcal{P}$ and $\mathcal{T}$ in our construction, {\it
by enforcing certain particle-hole symmetry of the fermionic
vortices.}

(3) As was pointed out in Ref.~\onlinecite{wanghall1,wanghall2},
the Wiedemann-Franz law $\kappa \sim T \sigma$ should generally be
strongly violated in a vortex liquid, as the vortices carry
entropy, but no charge. In the state we construct this is still
true, the modified Wiedmann-Franz law should be $ \kappa \sim L T
\sigma_v \sim L T \rho_e$. The Lorenz number $L$ is about
$\kappa/(T \sigma_e) \sim \rho_e^2$, which can be exceedingly
larger than an ordinary metal. Here we remind the readers that our
state has longitudinal transport only, and it can have both large
resistivity and thermal conductivity.

(4) For a strongly interacting electron system, the relaxation of
the electric current is pretty much independent from the
relaxation of a single particle. Hence the physics of a strongly
interacting electron liquid may be only captured by some
hydrodynamical description without microscopic
particles~\cite{delahydro,kovtunreview,hartnoll,hartnollreview,lucasreview,holoreview},
as hydrodynamics is defined at a much larger length scale. But
since the particle-vortex duality is a strong-weak duality, the
interaction between electron density becomes \beqn \sum_{i,j}
V_{i,j} n_i n_j \sim \int d^2x \ \frac{1}{g} (\vec{\nabla} \times
\vec{a})^2 \eeqn in the dual picture, where $g$ can be viewed as
the gauge coupling of the gauge charges (vortices), and also the
charge compressibility $\kappa_e$. The stronger the charge
interaction is, the weaker is the bare gauge coupling of the
vortices. \cxb{The common ``patch theory" for analyzing the RG
flow of a Fermi surface coupled with a U(1) gauge field predicts
that the gauge coupling would flow to a strongly coupled fixed
point eventually~\cite{nayak1,nayak2,sslee,maxnematic,mrossnfl}.
But this patch theory breaks down when there is disorder, as
disorder would mix different patches in the momentum space. But at
least when the bare gauge coupling $g$ is weak enough (which
corresponds to a strong charge density-density interaction), there
should be a sufficient window for the gauge coupling to be viewed
as a perturbation, and the momentum of the vortices can be
transferred to the photons, and then relax through disorder before
``feeding back" to the vortices.} Hence in this sense {\it we can
view the dual vortex system as an approximate vortex Fermi
liquid.}

\begin{center}
\begin{figure}
\includegraphics[width=0.4\textwidth]{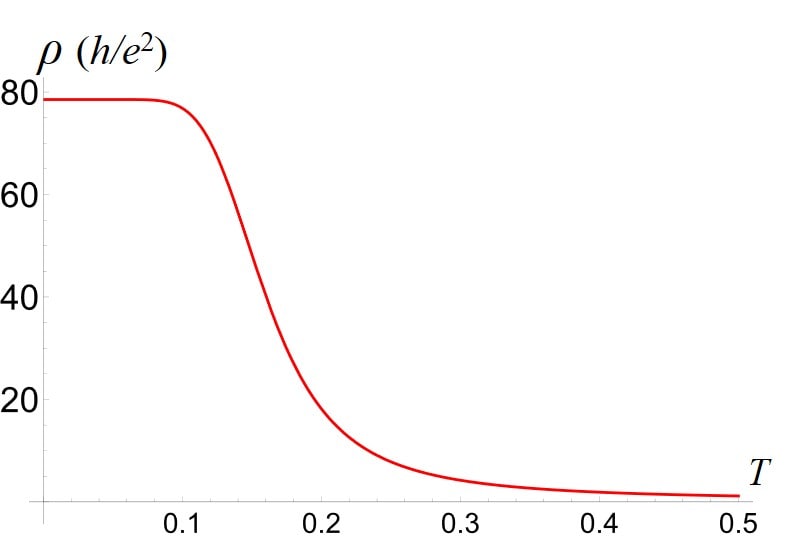}
\caption{The resistivity $\rho(T)$ as a function of temperature at
low temperature, computed using the composition formula
Eq.~\ref{composition}. We have chosen $\sigma_b = \exp(-
\Delta_b/T)$ where $\Delta_b = 1$, and $\sigma_f \sim 200$. $\rho$
is measured in unit of $h/e^2$ in this plot. We can also give
$\sigma_f$ a Fermi-liquid like temperature dependence, the plot
remains qualitatively unchanged.} \label{transport}
\end{figure}
\end{center}

\subsection{Quantum Bad metal at half-filling on a lattice}

The system we begin with is a strongly interacting electron system
with half-filling (one electron per site on average) on a lattice,
later we will discuss what happens when the system is doped away
from half-filling. We start with the standard ``$\SU(2)$ slave
rotor" theory for the electron
operator~\cite{su2gauge1,su2gauge2,su2rotor1,su2rotor2}: \beqn
c_{\ua} &=& f_{\ua} z_1 - f^\dagger_\da z^\dagger_2, \cr\cr
c_{\da} &=& f_{\da} z_1 + f^\dagger_{\ua} z^\dagger_2.
\label{parton}\eeqn Here $(f_\ua, f_{\da})$ is a fermionic spinon
doublet (fermionic partons) that carries spin-1/2, and $(z_1,
z_2)$ are slave rotors (bosonic partons) carrying the electric
charge. This formalism can maximally host a SU(2) charge
transformation and a SU(2) gauge transformation, and both
transformations can be made explicit by rewriting Eq.~\ref{parton}
in a matrix form (see $e.g.$ Ref.~\onlinecite{hermele2007,ransu2}
and references therein)~\footnote{An even more complete parton
construction can accommodate an SO(4) gauge
symmetry~\cite{xuso4}}. But for our purpose it suffices to assume
all SU(2) transformations, including the spin symmetry are broken
down to U(1). In fact, on a lattice with frustration, both the
charge SU(2) and the gauge SU(2) transformations are broken down
to U(1) by the most natural mean field states of $z_\alpha$ and
$f_\alpha$. The assignment of the electric charge symmetry
$\U(1)_e$, gauge symmetry $\U(1)_g$ and spin symmetry $\U(1)_s$ on
the partons is \beqn \U(1)_e &:& z_1 \ra e^{\ii \theta_e} z_1, \ \
\ z_2 \ra e^{- \ii \theta_e} z_2, \ \ \ f_\alpha \ra f_\alpha;
\cr\cr \U(1)_g &:& z_1 \ra e^{\ii \theta_g} z_1, \ \ \ z_2 \ra
e^{\ii \theta_g} z_2, \ \ \ f_\alpha \ra e^{- \ii \theta_g}
f_\alpha, \cr\cr \U(1)_s &:& z_\alpha \ra z_\alpha, \ \ \ f_\ua
\ra e^{\ii \theta_s} f_\ua, \ \ \ f_\da \ra e^{- \ii \theta_s}
f_\da. \eeqn The system being at half-filling implies that the
total rotor number of $z_1$ and $z_2$ are equal: $\sum_i n_{1,i} =
\sum_i n_{2,i}$. There is a dynamical $\U(1)_g$ gauge field
$a_\mu$ that couples to both $z_\alpha$ and $f_\alpha$. The U(1)
gauge constraint demands that on every site $i$, $\sum_\alpha
n_{\alpha,i} + 1 = \sum_\alpha f^\dagger_{\alpha,i} f_{\alpha,i}$.
A detailed discussion of the physical meaning of the partons
introduced can be found in Ref.~\onlinecite{ransu2}.

This slave rotor parton construction allows us to construct many
states of the strongly interacting electron system which are
difficult to visualize using free or weakly interacting electrons.
For example, if the bosonic partons $z_\alpha$ are in a trivial
bosonic Mott insulator state with $n_{1,i} = n_{2,i} = 0$ (meaning
$\sum_\alpha f^\dagger_{\alpha,i} f_{\alpha,i} = 1$ on each site),
the system becomes a Mott insulator of electrons with a charge
gap, and the spin physics of the Mott insulator depends on the
state of $f_\alpha$. Various spin liquid states can be designed
and classified depending on the mean field band structure of
$f_\alpha$~\cite{wen2002}.


The many-body state of electrons is determined by the states of
the partons. In the last decade the study of symmetry protected
topological states (SPT) significantly broadened our understanding
of the states of matter~\cite{wenspt,wenspt2}, which also allows
us to construct even more novel states of electrons using the
partons. We first use $z_\alpha$ to define two other composite
bosonic fields: $\phi_e = z^\dagger_1 z_2$, $\phi_g = z_1z_2$.
$\phi_e$ and $\phi_g$ carry charge $(2, 0)$ and $(0, 2)$ under
$\left( \U(1)_e, \U(1)_g \right)$. Then we drive the composite
fields $\phi_e$ and $\phi_g$ into a bosonic SPT state with
$\U(1)_e$ and $\U(1)_g$ symmetries, which is the bSPT state for
two flavors of bosons constructed in
Ref.~\onlinecite{levinsenthil}. The physics of this bSPT state is
analogous to the quantum spin Hall insulator: the vortex of
$\phi_e$ carries charge of $\phi_g$, and vice versa. If we follow
the Chern-Simons description of the bSPT~\cite{luashvin}, this
state is \beqn \mathcal{L}_{\mathrm{bSPT}} = \frac{\ii
K^{IJ}}{2\pi} \tilde{a}_I \wedge d \tilde{a}_J + \frac{\ii
2}{2\pi} \tilde{a}_1 \wedge da + \frac{\ii 2}{2\pi}
\tilde{a}_2\wedge dA^e, \label{sptcs}\eeqn where $K^{IJ}$ takes
the same form as the Pauli matrix $\sigma^x$. Here $\ast
d\tilde{a}_1$ and $\ast d\tilde{a}_2$ are the dual of the currents
of $\phi_g$ and $\phi_e$ respectively. \cxc{This bSPT state of the
rotors also has a particle-hole symmetry of the bosonic rotors,
and in this bSPT state the expectation value of the total rotor
number of both $z_1$ and $z_2$ is zero. Please note that the total
rotor number being zero does not imply a trivial vacuum state, as
the rotor number (just like a spin $S^z$ operator) can take both
positive and negative values. }

A bSPT state is gapped, and also nondegenerate, hence it is safe
to integrate out the bosonic degree of freedom, and obtain the
response to the gauge fields. After integrating out
$\tilde{a}_{1,2}$ from Eq.~\ref{sptcs}, we obtain: \beqn
\mathcal{L} &=& \mathcal{L}_{F}(f_\alpha, a_\mu) +
\frac{4\ii}{2\pi} a \wedge dA^e + \frac{1}{g} (\vec{\nabla} \times
\vec{a})^2 + \cdots. \label{cs} \eeqn One can also introduce the
external gauge field for the spin symmetry $\U(1)_s$, but it won't
have a nontrivial response from the bSPT state.

The mutual Chern-Simons term in the last term of Eq.~\ref{cs}
fundamentally changes the physics of the system in the following
way:

The electric charge current, which is defined as $J^e = \delta
\mathcal{L} / \delta A_e$, is identified as \beqn J^e = e
\frac{4}{2\pi} \ast d a ,\eeqn meaning the flux of $a_\mu$ now
carries electric charge $4e$. Hence the bSPT state so constructed
turns the gauge field $a_\mu$ into the dual of the charge current
in the sense of the particle-vortex
duality~\cite{peskindual,halperindual,leedual}, and turns the
gauge charge of $a_\mu$ into the charge vortex. {\it The fermionic
parton $f_\alpha$, which carries gauge charge $1$ under $a_\mu$,
now automatically becomes the vortex of the electric charge}, as
when a charge (now the flux of $a_\mu$) circulates the gauge
charged $f_\alpha$, it would accumulate a Berry's phase. It is
worth noting that,  for more general bSPT states of $z_\alpha$
with only mutual $\U(1)_e\times \U(1)_g$ Chern-Simons response,
the electric charge carried by the flux of $a_\mu$ has to be an
integer multiple of $4e$. Hence, the bSPT described above is the
minimal non-trivial choice.

Now we take the long wavelength limit, and integrate out both
$f_\alpha$ and $a_\mu$; we also choose the temporal gauge with
$a_\tau = 0$. The response Lagrangian in terms of $A^e$ is \beqn
\mathcal{L}_{\mathrm{res}} = \sum_{\omega} \frac{1}{2}
\frac{4}{\pi^2}\frac{\omega^2}{\Pi_f(\omega)}|\vec{A}^{e,
t}(\omega)|^2. \eeqn $\vec{A}^{e,t}$ is the transverse component
of $\vec{A}^e$. $\Pi_f(\omega)$ is the polarization of $f_\alpha$,
and it should be proportional to $\ii \omega \sigma_f(\omega)$
after analytic continuation~\cite{NagaosaLee}, where
$\sigma_f(\omega)$ is the conductivity of the fermionic parton
($i.e.$ also the vortex) $f_\alpha$. This implies that the
electrical resistivity of the system should be \beqn
\rho_e(\omega) = \frac{\pi^2}{4} \sigma_f(\omega) =
\frac{\pi^2}{4} \left( \frac{\sigma_0}{1 - \ii \omega \tau_v}
\right). \label{res}\eeqn We note that here $\rho_e$ is measured
with unit $\hbar/e^2$; $\sigma_f$ is computed in the convention
that $f_\alpha$ carries charge 1. Here we can evaluate the
conductivity of the good metal of $f_\alpha$ using the rudimentary
Drude formula. $\sigma_0$ is the conductivity of the $f_\alpha$ in
the DC limit. It was shown recently that even when a Fermi surface
is coupled to a dynamical gauge field, the response to the gauge
field is still exactly the same as what is computed by the Drude
theory (at least when there is no disorder)~\cite{drudeanomaly}.
\cxb{We also exploit the fact that, when the electron density has
a strong interaction, the bare gauge coupling becomes weak, and
the photon-vortex interaction remains perturbative at least within
a large window of scale.} An analysis of the fermions interacting
with gauge field in a disordered environment can be found in
Ref.~\onlinecite{srimulligan}.

In Eq.\ref{res}, $\sigma_0$ can be rather large, namely the
vortices form a good metal, when there is a finite Fermi surface
of $f_\alpha$ and the disorder is weak. In this case the
electrical resistivity of the system can be far beyond the MIR
limit, $i.e.$ the system is a very bad metal.

Now we investigate the spatial reflection symmetry $\mathcal{P}$
of our system. And let us use $\mathcal{P}_y: y \ra -y$ as an
example. We assume that $c_\alpha$ changes up to a sign under
$\mathcal{P}_y$, then this leads to the transformation of
$f_\alpha$, $z_\alpha$: \beqn \mathcal{P}_y: ~~ f_\alpha \ra
(\sigma^x)_{\alpha\beta} f_\alpha^\dagger, \ \ \ z_\alpha \ra
(\sigma^x)_{\alpha\beta} z^\dagger_\beta. \eeqn The $\U(1)_e$ and
$\U(1)_g$ charges are even and odd under $\mathcal{P}_y$
respectively. This means that $\tilde{a}_{1,2}$ transform
oppositely under $\mathcal{P}$, and the bSPT state preserves
$\mathcal{P}$ based on Eq.~\ref{sptcs}. In order to ensure the
reflection symmetry, we also need the band structure of $f_\alpha$
to satisfy $ \varepsilon_\ua(k_x, k_y) = - \varepsilon_\da(k_x,
-k_y)$.

Our construction also preserves a (special) time-reversal symmetry
$\mathcal{T}$ defined as following: the electron operator is still
invariant under $\mathcal{T}$ (up to an extra sign). We can choose
the following transformations of $z_\alpha$ and $f_\alpha$ to
ensure the desired transformation of the electrons: \beqn
\mathcal{T}: ~~ z_\alpha \ra (\sigma^x)_{\alpha\beta}
z^\dagger_{\beta}, \ \ \ f_\alpha \ra (\sigma^x)_{\alpha\beta}
f^\dagger_\beta. \eeqn As we can see, the $\U(1)_e$ and $\U(1)_g$
charges are again even and odd under $\mathcal{T}$ respectively.
This means that $\tilde{a}_{1,2}$ transform oppositely under
$\mathcal{T}$, and the bSPT state preserves $\mathcal{T}$ based on
Eq.~\ref{sptcs}. In order to ensure the time-reversal symmetry, we
also need the band structure of $f_\alpha$ to satisfy $
\varepsilon_\ua(\vec{k}) = - \varepsilon_\da(- \vec{k})$. More
precisely, here the time-reversal is a product between the
particle-hole transformation and a
spatial-inversion~\footnote{\cxd{Without the spatial-inversion
transformation, the time-reversal transformation alone would
demand $\varepsilon_\ua(\vec{k}) = - \varepsilon_\da(\vec{k})$,
which would make the Fermi pockets of the two spinon bands overlap
with each other, and lead to a potential exciton instablity due to
nesting.} }. A more realistic time-reversal symmetry for electrons
with $\mathcal{T}^2 = -1$ can be defined and preserved if we
introduce another orbital flavor to the electrons.

So far we have ignored the conductivity of the bosonic partons,
which is valid when the energy scale is much smaller than the gap
of the bosons $\Delta_b$. With finite frequency $\omega$, the
bosons will also make two new nonzero contributions to the
longitudinal response of $A^e$. The first of which is a simple
addition to the response Lagrangian of the boson polarization
$i.e.$ $\mathcal{L}^b_{\mathrm{res}} \sim
\frac{1}{2}\Pi_b(\omega)|\vec{A}^{e,t}(\omega)|^2 $, for which
$\Pi_b(\omega) \rightarrow \ii \omega \sigma_b(\omega)$ after
analytic continuation. This only modifies the charge conductivity
by shifting it a value of $\sigma_b(\omega)$. The second, more
interesting contribution, is that a longitudinal term will be
generated for the internal gauge field $a_\mu$ as well. In
principle $\Pi_b(\omega)$ can be different for $A^e_\mu$ and
$a_\mu$, but without loss of generality we assume that they are
the same. Since the internal gauge field and electromagnetic gauge
field transform differently under $\mcal{P}$ and $\mcal{T}$ and so
do the $\phi_g$ and $\phi_e$ current, there cannot be mixed terms
between $a$ and $A^e$ that would lead to mutual longitudinal
response in the effective Lagrangian. Then eventually the
electrical conductivity of the system follows the following
composition rule \beqn \sigma_e = \sigma_b + \frac{4}{\pi^2}
\left( \frac{1}{\sigma_f + \sigma_b} \right). \label{composition}
\eeqn In this equation, $\sigma_e$ is measured in the unit of
$e^2/\hbar$; $\sigma_b$ and $\sigma_f$ are computed in the
convention of charge $1$ and $\hbar = 1$. This composition is very
different from the Ioffe-Larkin rule~\cite{IoffeLarkin}, and we
expect this composition rule to be valid for temperature far
smaller than the gap of the bosonic parton, $i.e.$ when the
thermally activated bosonic partons are very dilute. In
Fig.~\ref{transport} we plot $\rho_e(T) = 1/\sigma_e(T)$ measured
in unit of $h/e^2$ using the composition rule
Eq.~\ref{composition}.

Another quantity of interest is the charge compressibility. The
compressibility can be computed from the charge density-density
correlation function which can be attained by reading off the
$A_\tau(q) A_\tau(-q)$ term in $\mcal{L}_{\text{res}}$ at
vanishing frequency after integrating out all the matter and the
internal gauge fields. As we discussed before, one contribution to
the compressibility is proportional to the gauge coupling $g$,
which is evident after integrating out $\vec{a}$ from
Eq.~\ref{cs}. Eventually the compressibility should involve the
bare charge density-density interaction, after being renormalized
by integrating out the fermions $f_\alpha$. In the limit $\omega
\rightarrow 0$ and large gap of bosonic rotors, the total
compressibility is given by \beqn \kappa_e(\vb{q}) = \frac{1}{
(\kappa_0)^{-1} + \frac{\pi^2}{4} \frac{\Pi_f(q)}{|\vb{q}|^2} },
\eeqn where $\kappa_0$ is the ``bare" compressibility of the
system when the gauge field $a_\mu$ is not coupled to any matter
fields, hence $\kappa_0$ is proportional to the gauge coupling
$g$: $\kappa_0 = g (4/\pi^2)$. If we choose a simple quadratic
dispersion or a circular Fermi surface for the vortices, then the
result for the fermionic vortex polarization at zero frequency is
well-known to be~\cite{PhysRevB.76.165104,NagaosaLee}
$\Pi_f(\vb{q}) \sim |\vb{q}|^2/(12 \pi m_v)$ where $m_v$ is the
effective mass of the fermionic vortices. Like the charge
conductivity, this gives us a composition rule for the charge
compressibility in terms of linear response functions for the
different species of partons. For the more general case of a
finite boson gap, this composition rule involves both the boson
compressibility and boson polarization $\Pi_b \sim \chi_b
|\vb{q}|^2$: \beqn \kappa_e(\vb{q}) = \kappa_b(\vb{q}) +
 \frac{1}{(\kappa_0)^{-1} + (\chi_b + \chi_f)\pi^2/4}. \eeqn
Here, $\chi_b$ is the magnetic susceptibility of the bosons. We
expect this composition rule to hold at small finite temperature
much below the boson gap.

\subsection{Physics at weak doping}

\begin{center}
\begin{figure}
\includegraphics[width=0.4\textwidth]{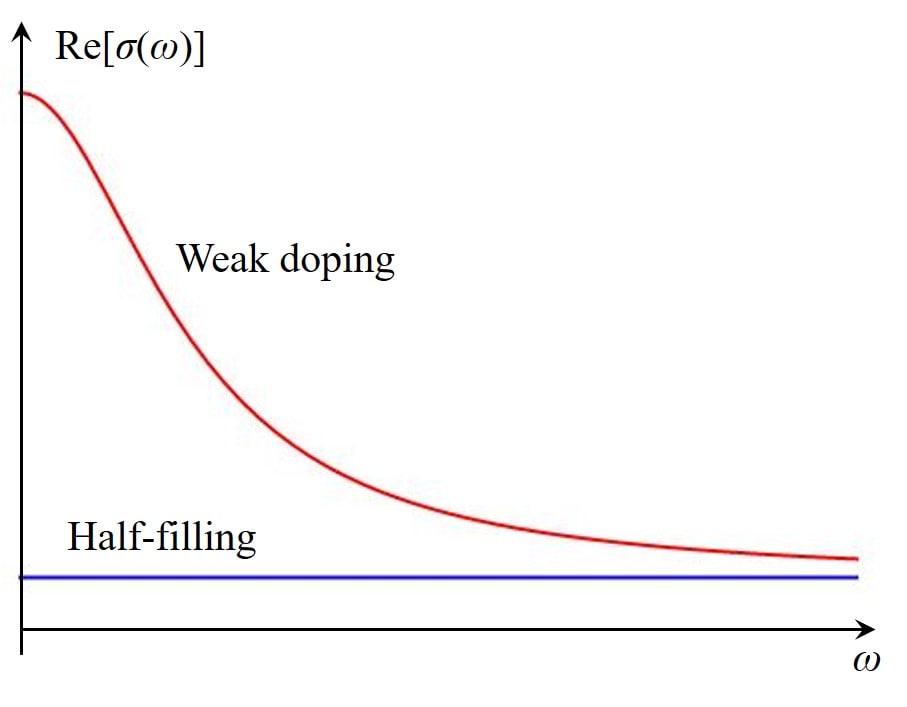}
\caption{The optical conductivity $\mathrm{Re}[\sigma(\omega)]$
near $\omega = 0$ at half-filling, and with small doping $\delta
n_e$. } \label{opticalcond}
\end{figure}
\end{center}

We would also like to consider the effects of weak charge doping
$\delta n_e$ away from one electron per site. As the charge
density is now bound with the internal gauge flux through the
mutual CS term in Eq.~\ref{cs}, weakly doping the system
corresponds to adding a background $\U(1)_g$ gauge field with
small average magnetic flux. Since the vortices (spinon
$f_\alpha$) form a good metal, their conductivity may still be
evaluated through the semiclassical Drude theory. Using the
semiclassical equation of motion with magnetic field, it is
straight forward to compute the Drude conductivity of $f_\alpha$:
\beqn \sigma_f(\omega) = \bigg ( \frac{\frac{1}{\tau_v} - i
\omega}{\omega_c^2 + ( \frac{1}{\tau_v} - i \omega)^2} \bigg )
\frac{1}{\tau_v}\sigma_{0}. \eeqn $\omega_c$ is the cyclotron
frequency of the vortices $\omega_c = \bar{b}/m_v$, and $\bar{b}
\sim \delta n_e$ is the average flux seen by the fermionic
vortices. Note that the conductivity $\sigma_f(\omega)$ of
$f_\alpha$ only contains a longitudinal component given by the
expression above.  The Hall conductivities from $f_\uparrow$ and
$f_\downarrow$ cancel each other due to the time-reversal symmetry
$ \mathcal{T}$ which involves a particle-hole transformation of
the spinons $f_\alpha$. From the relation between the charge
conductivity and vortex conductivity in Eq.~\ref{composition}, if
the boson gap is taken to infinity, we can extract the main
interesting piece of the charge conductivity, which is given by
\beqn \sigma_e(\omega) \sim \frac{4}{\pi^2 \sigma_f(\omega)} =
\frac{4 \tau_v}{ \pi^2 \sigma_{0}} \bigg
[\frac{\omega_c^2}{\frac{1}{\tau_v} - i \omega} + \bigg
(\frac{1}{\tau_v} - i \omega  \bigg ) \bigg ]. \eeqn At zero
doping, the optical conductivity does not have a Drude weight;
{\it but the presence of added charge has introduced a new Drude
peak to the optical conductivity with Drude weight:} \beqn D \sim
\frac{2 \omega_c^2 \tau_v}{\pi^2 \sigma_{0}} \sim (\delta n_e)^2,
\eeqn and the Drude weight is proportional the square of the doped
charge density, contrary to the ordinary Drude theory where the
Drude weight is linear with the charge density.

The DC resistivity now takes the form of \beqn \rho_e =
1/\sigma_e(0) = \frac{\pi^2 \sigma_0}{4} \frac{1}{1 + \tau_v^2
\omega_c^2}. \eeqn The Lorenz number, defined as $L = \kappa/(T
\sigma_e)$ becomes \beqn L = \frac{\kappa}{T \sigma_e} \sim
\frac{L_0 \sigma_f}{\sigma_e} \sim \frac{\pi^2}{4}\frac{L_0
\sigma_0^2}{(1 + \tau_v^2 \omega_c^2)^2}. \eeqn \cxb{Here $\kappa$
represents the thermal conductivity.} Both the resistivity, and
Lorenz number decrease with the doped charge density. Combing with
the emergence of the Drude weight under doping, it suggests that
doping would eventually drive the the system more like a normal
metal. Here we have ignored the thermal conductivity arising from
the gauge bosons, which also transports heat without charge, hence
it also contributes to the violation of the Wiedemann-Franz law.

\subsection{Nearby Phases}

(1) {\it Spin liquid Mott insulator, and metal}

The quantum bad metal state constructed above can be driven to a
Mott insulator which is also a $\U(1)$ spin liquid with a Fermi
surface of spinon $f_\alpha$, by driven $(z_1, z_2)$ into a
trivial bosonic Mott insulator, with zero rotor number of $z_1$
and $z_2$. In the Mott insulator, $f_\alpha$ still has Fermi
pockets, but the gauge flux of $a_\mu$ no longer carries any
nontrivial quantum number. This is one of the most studied spin
liquid states in the
literature~\cite{lee2005,motrunich2005,ran2007}, with potential
applications to a variety of materials.

The quantum phase transition between a bSPT state and a trivial
Mott insulator of the boson is described by the $N=2$
QED~\cite{groverashvin,luashvin}, and this theory is part of a web
of duality involving also the easy-plane deconfined quantum
critical
point~\cite{xudual,mrossdual,seiberg2,potterdual,SO5,dualreview}.
The original theory of the bSPT-MI transition is now coupled to an
extra dynamical gauge field $a_\mu$. When there is no disorder,
the dynamics of $a_\mu$ is overdamped by the Fermi surface of
$f_\alpha$, then we do not expect the gauge field $a_\mu$ to
change the infrared fate of the transition. The presence of
disorder may complicate the nature of the bSPT-MI transition.

The quantum bad metal phase can also be driven into an ordinary
metal phase by condensing either $z_1$ or $z_2$. The $\U(1)_g$
gauge field will be gapped out by the Higgs mechanism, and the
spinon operator $f_\alpha$ becomes the electron operator due to
the condensate of, $e.g.$ $z_1$, similar to the previous theory of
interaction-driven MIT~\cite{lee2005,senthilMIT}.

(2) {\it Charge-$4e$ superconductor}

Starting with the quantum bad metal, we can also drive the spinon
$f_\alpha$ into a trivial insulator without special topological
response, likely through a Lifshitz transition where the Fermi
pockets shrink to zero.
Then the action of $a_\mu$ is just
the ordinary Maxwell term, which describes a photon phase. The
monopole which creates and annihilates the gauge flux is
prohibited here as the gauge flux carries charge-$4e$ as we
discussed before, and the electric charge is a conserved quantity.
The photon phase of the gauge field $a_\mu$ is also dual to the
condensate of its flux, $i.e.$ a condensate of charge$-4e$, or in
other words a charge$-4e$ superconductor.

(3) {\it $Z_2$ spin-charge topological order}

We can also consider a situation where the fermionic vortices
$f_\alpha$ form a ``superconductor", $i.e.$ the Cooper pair of
$f_\alpha$ condenses. This condensation will gap out $a_\mu$
through a Higgs mechanism, and break $a_\mu$ down to a $Z_2$ gauge
field, which supposedly forms a $Z_2$ topological order. Like all
the $Z_2$ topological orders, here there are three types of anyons
with mutual semionic statistics. One type of anyon is $f_\alpha$,
another is the half-flux of $a_\mu$. Since the flux of $a_\mu$
carries charge-$4e$, the half-flux of $a_\mu$ carries charge-$2e$.

\subsection{Other Constructions}

One can also construct a similar quantum bad metal phase starting
with a charge-$2e$ spin-singlet superconductor. Let us assume
there are two flavors of bosons, $b_1$ and $b_2$, which carry
charge $\pm 2e$ respectively. We take the following parton
treatment for $b_\alpha$: \beqn b_1 = \psi_1 f, \ \ \ b_2 = \psi_2
f, \eeqn where all the partons are complex fermions. Apparently
there is also a gauge $U(1)_g$ shared by the partons. $\psi_{1,2}$
carries electric charge $\pm 2$, and gauge charge $+1$ of a
dynamical internal gauge field; $f$ carries gauge charge $-1$ of
the internal gauge field. We now consider the following state: $f$
again forms a band structure which is a good metal; $\psi_\alpha$
forms a quantum ``psudospin" Hall insulator, in the sense that the
flavor index of $\psi_\alpha$ is viewed as a pseudospin index.
After integrating out $\psi_\alpha$, a mutual Chern-Simons term
between the external EM field $A^e$ and the internal gauge field
$a$ is generated, with the same form as Eq.~\ref{cs}. We assume
that there is no other conserved charges other than the charge
$\U(1)$. The charge response of this construction can be evaluated
following the steps of the previous section.

The bSPT state is one of the states that the bosonic partons
$(z_1, z_2)$ can form that make the charge vortex a fermion. There
are other options which achieve the similar effect, if we allow
topological degeneracy. For example, $z_1$ and $z_2$ can each form
a bosonic fractional quantum Hall state with Hall conductivity
$\pm 1/(2k)$ where $k$ is an integer. Although there is a bit
subtlety of integrating out a topological order, suppose we can do
this, the response mutual CS theory in Eq.~\ref{cs} would have
level $2/k$. The rest of the discussion follows directly.

We would like to compare our state with other ``vortex liquids"
discussed in previous literature, for example the well-known Dirac
vortex liquid in the context of half-filled Landau
level~\cite{son2015,wanghall2}. The electrical conductivity tensor
of the system reads $ \sigma_{ij} = \delta_{ij} \sigma^\ast +
\frac{e^2}{2h}\epsilon_{ij} $, where $\sigma^\ast \sim
1/\sigma_v$, and $\sigma_v$ is the conductivity of the Dirac
composite fermions (the vortices). Although the longitudinal
conductivity of the system can be small when $\sigma_v$ is large,
the longitudinal resistivity would still be small due to the
nonzero Hall conductivity. Hence in the simplest experimental
set-up where the transport is measured along the $\hat{x}$
direction while the $\hat{y}$ direction of the sample has an open
boundary, the measured longitudinal resistivity along the $x$
direction would be small.

We can also put $c_\ua$ in the Dirac vortex liquid of the
half-filled Landau level; and $c_\da$ in the time-reversal
conjugate state of $c_\ua$. In this case in order to correctly
extract the longitudinal electrical resistivity, we need to
introduce both the external electromagnetic field, and a ``spin
gauge field" $A^s$: \beqn \mathcal{L} &=& \mathcal{L}_D(\psi_1,
a_1) + \frac{\ii}{4\pi} a_1 \wedge A_1 + \frac{\ii}{8\pi} A_1
\wedge dA_1, \cr\cr &+& \mathcal{L}_D(\psi_2, a_2)
-\frac{\ii}{4\pi} a_2 \wedge A_2  - \frac{\ii}{8\pi} A_2 \wedge
dA_2. \eeqn $\psi_{1,2}$ are the composite Dirac fermions, and
their Lagrangian $\mathcal{L}_D$ should in general have a nonzero
chemical potential. We have introduced two external gauge fields
$A_1 = A^e + A^s$, and $A_2 = A^e - A^s$. The system does not have
a net charge Hall response, but there is a spin-Hall effect,
$i.e.$ there is a mutual CS term between the electromagnetic field
and the spin gauge field. The existence of the spin Hall response
will again lead to a small longitudinal electrical resistivity
when $\psi_{1,2}$ are good metals. By contrast, in our
construction presented in the previous section, there is only
longitudinal transport, hence a large vortex conductivity would
ensure a large longitudinal electrical resistivity.

Another vortex liquid with fermionic vortex was discussed in
Ref.~\onlinecite{vortexliquidfisher}, aiming to understand the
observed metallic state with an anomalously large conductivity
between the superconductor and insulator in amorphous thin films.
There the vortex is turned into a fermion through manual flux
attachment.

Vortices can also play an important role in quantum magnets.
Exotic quantum spin liquid states were also constructed through
fermionic spin vortices in previous
literature~\cite{spinvortex1,spinvortex2,spinvortex3,spinvortex4,spinvortex5}.
These works generally use two approaches to generate fermionic
vortices: one can either introduce fermionic partons for the
vortex operator, or turn a vortex into a fermion through flux
attachment when the vortex sees a background magnetic field (dual
of fractional spin density). In our work, instead of granting
existing vortices fermionic statistics, an interpretation of the
fermionic partons as charge vortices naturally emerges due to the
topological physics of the bosonic sector.

\section{Summary}

We present a construction of a strongly interacting quantum bad
metal phase, $i.e.$ at zero and low temperature the resistivity is
finite but exceedingly larger than the MIT limit $h/e^2$, by
making the charge vortices a good metallic phase with a vortex
Fermi surface. In this construction the charge vortex is naturally
a fermion by driving the charge degree of freedom into a bosonic
symmetry protected topological state. The quantum bad metal so
constructed has the following features: (1) its resistivity can be
exceedingly larger than the MIR limit; (2) a small Drude weight
proportional to $(\delta n_e)^2$ emerges under weak charge doping
away from half-filling (one electron per unit cell); (3) like
previously discussed vortex liquids, our construction should also
have strong violation of the Wiedemann-Franz law. We also
demonstrated that this quantum bad metal phase is next to a
charge-$4e$ superconductor, a $Z_2$ spin-charge topological order,
the Mott insulator phase which is also a well-studied spin liquid,
and a normal metal phase.

The exoticness of the state we constructed is of ``quantum
nature", as strictly speaking a bSPT state that our construction
strongly relies on is only sharply defined at zero temperature. At
high temperature all the partons will be confined, and our state
no longer enjoys a tractable description in terms of the dual
weakly interacting vortices. Hence our state is different from the
original example of bad metal discussed in
Ref.~\onlinecite{bad_metal}, $i.e.$ the cuprates materials with
hole doping, where the resistivity of the system in each $2d$
layer reaches the threshold of bad metal at finite temperature.

The authors thank Matthew Fisher and Steve Kivelson for very
helpful discussions. We also thank Umang Mehta and Xiao-Chuan Wu
for participating in the early stage of the work. This work is
supported by the NSF Grant No. DMR-1920434, and the Simons
Investigator program. C.-M. J. is supported by a faculty startup
grant at Cornell University

\bibliography{vortexFS}

\end{document}